\documentclass[twocolumn,aps,pra]{revtex4}
\usepackage{epsfig}
\usepackage[english]{babel}
\usepackage{latexsym}
\usepackage{graphics}
\usepackage{subfigure}
\usepackage{graphicx}
\usepackage{dcolumn}
\usepackage{amsmath}
\usepackage{hyperref}
\usepackage{amssymb}
\usepackage{color}
\usepackage{setspace}

\begin{document}
\title{Response time of electrons to light in strong-field ionization of polar molecules}
\author{J. Y. Che$^{1}$, S. Ye$^{2}$, S. Q. Shen$^{2}$, W. Y. Li$^{3,*}$ and Y. J. Chen$^{2}$}

\date{\today}

\begin{abstract}
We study tunneling ionization of HeH$^+$ in strong elliptical laser fields numerically and analytically. The calculated photoelectron momentum distribution (PMD) show two different offset angles corresponding to ionization events occurring in the first and the second half cycles of one laser cycle. When the larger angle is greater than the angle of a model symmetric molecule with a similar ionization potential to the polar molecule, the smaller angle is smaller than the symmetric molecule. Using a developed strong-field model that consider effects of both the permanent dipole (PD) and the asymmetric Coulomb potential, we are able to quantitatively reproduce these phenomena. We show that the PD effect can increase (decrease) the response time of electrons within polar molecules to light in photoemission, thereby increasing (decreasing) the offset angle related to the first (second) half cycle of a laser cycle. The laser-dressed asymmetric Coulomb potential near the atomic nuclei also plays an important role in the sub-cycle-related response time and offset angle. This model may be useful for quantitatively studying attosecond ionization dynamics of polar molecules in strong laser fields.  

\end{abstract}

\affiliation{1.School of Physics, Henan Normal University, Xinxiang, China\\ 2.College of Physics and Information Technology and Quantum Materials and Devices Key Laboratory of Shaanxi Province's High Education Institution, Shaan'xi Normal University, Xi'an, China\\ 3.Hebei Key Laboratory of Optoelectronic Information and Geo-detection Technology, School of Mathematics and Science, Hebei GEO University, Shijiazhuang, China}
\maketitle

\section{Introduction}
In the past two decades, attosecond science has provided unprecedented access for detecting ultrafast processes in atoms \cite{McPherson,Schafer2,Tong1997}, molecules \cite{CorkumNP2007}, and solids \cite{TaoZ2016}, allowing people to measure microstructure and control electronic dynamics with a time-space resolution on the atomic scale \cite{NeutzeR2000,HentschelM2001,UdemT2003}. The development of attosecond science is closely related to strong-field physics. For the system of an atom or a molecule interacting with a strong laser field, the bound electron can escape from the laser-Coulomb-formed barrier through tunneling \cite{TaoZ2016,NeutzeR2000}. This tunneling event triggers rich physical processes such as above-threshold ionization (ATI) \cite{Schafer1,Yang1993,Lewenstein1995,Paulus2002}, high-order harmonic generation (HHG) \cite{Corkum,Lewenstein1994}, and non-sequential double ionization (NSDI) \cite{Ivanov1,Corkum2,Becker1}, etc.. 
These processes have important applications in attosecond science \cite{Krausz2009,Vrakking2014}. Current attosecond measurements are not direct. After obtaining experimental results, one needs to use theoretical models to invert the time information of intermediate processes \cite{Eckle2008,Ivanov2012}. Therefore, the accuracy of the theoretical model will directly influence the accuracy of the inversion results \cite{xie2020}. Furthermore, from an application perspective, it is also highly desirable to establish an analytical theory that can provide a simple and one-to-one mapping between experimental observations and time-domain information of ultrafast processes, making it easy to extract temporal information from experimental measurements.

Theory models of strong-field approximation (SFA) \cite{Lewenstein1995,Lewenstein1994} and Coulomb-corrected SFA \cite{MishaY, Goreslavski, yantm2010} have been widely used in strong-field physics. The SFA ignores the Coulomb effect, but can provide a simple mapping between time and observable values. The Coulomb-corrected SFA takes into account the Coulomb effect through solutions of Newton equations including both electric-field force and Coulomb force, but is not convenient to provide an analytical relationship between time and observable values. Recently, a semiclassical analytical theory is developed to describe the response time of the electron inside an atom to light in strong-field tunneling ionization \cite{chen2021,che2023,cheaps}. This time reflects the essential response of the electronic wave function to a tunneling ionization event during photoemission, and characterizes the strong three-body interaction time between Coulomb, electron, and external field. The observable values inferred from the response time predicted by this theory \cite{chen2021,che2023,cheaps} are consistent with recent attoclock experiments for different laser and atomic parameters \cite{Sainadh2019,Torlina2015,Pfeiffer2012,Boge2013,Quan2019}, providing other insights into the tunneling time \cite{Ramos2020,Li2022}. Especially, this theory can analytically provide scaling laws for the dependence of experimental observations on laser parameters \cite{che2023}. It can also be applied to diverse forms of laser field such as orthogonal two-color (OTC) laser fields \cite{Wu2023} and symmetric molecules \cite{gen2022,shen2024}. With further considering the Coulomb acceleration \cite{Lein2002,Itatani2004,chen2009} in the rescattering process, this theory can also be generalized to quantitatively describe more complex strong-field processes such as HHG \cite{peng2025}. However, compared to atoms and symmetric molecules, polar molecules show more complex responses to the laser field due to structural asymmetry. For example, due to the effects of permanent dipole (PD) arising from symmetry breaking, asymmetric ionization related to laser-dressed ground-state energy occurs in each laser cycle \cite{KamtaGL2005}. The combined effects of the PD and Coulomb potential also lead to the significant asymmetry of PMDs, when the polar molecule is exposed to the linearly polarized single-color laser field \cite{WS2020}. Different ionization channels related to the ground state and the first excited state of the asymmetric system and subtle sub-cycle electron dynamics information can also be mapped in different quadrants of PMDs in OTC fields \cite{Che2021pra}, etc.. For the reasons, there is currently no analytical theory that can quantitatively describe the intrinsic correlation between response time and experimental observations in strong-field ionization of polar molecules.

In this paper, we focus on the response time of electrons inside polar molecules to strong low-frequency laser fields and make efforts to construct an analytical response-time theory for laser-induced tunneling ionization of polar molecules. Through numerical solution of time-dependent Schr\"{o}dinger equation (TDSE), we study the ionization of oriented polar molecules in strong elliptically-polarized laser fields with high ellipticity. The calculated PMD offset angle of polar molecules in the first half cycle is larger than that in the second half cycle, and the angle difference between the adjacent laser sub-cycles increases with the increase of internuclear distance. In comparison with model symmetric molecules and atoms with similar ionization potential to polar molecules, the larger angle of polar molecule is greater than that of symmetric molecule and the smaller one is smaller than the symmetric molecule but is larger than that of atom. With considering the effects of PD and Coulomb potential, we generalize the semiclassical response-time theory to polar molecules. The generalized theory model is able to give an analytical expression for PD-related response time and well reproduce the half-cycle-related offset angle of TDSE simulations. This model reveals that the PD effect can increase (decrease) the response time of electrons inside polar molecules to strong lasers in tunneling ionization and therefore increase (decrease) the offset angle in the first (second) half cycle of one laser cycle. Furthermore, the characteristics of the asymmetric two-center Coulomb potential dressed by laser near the atomic nuclei also influence remarkably on the sub-cycle ionization dynamics of the polar molecule. This model may be useful for precisely probing attosecond dynamics of oriented polar molecules in strong-field ionization.

\section{Method}
\subsection{Numerical methods}
In length gauge, the Hamiltonian of the atom or molecule system interacting with a strong laser field can be written as (in atomic units of $ \hbar = e = m_{e} = 1$)
\begin{equation}
{H}(t)= H_{0}+\mathbf{E}(t)\cdot \mathbf{r}
\end{equation}
Here, $H_{0}={\mathbf{{p}}^2}/{2}+V(\mathbf{r})$ is the field-free Hamiltonian and $V(\textbf{r})$ is the Coulomb potential of the system.
The potential $V(\textbf{r})$ used for molecule has the form of
$V(\mathbf{r})=-Z_1/\sqrt{\eta+\mathbf{r}_1^{2}}-Z_2/\sqrt{\eta+\mathbf{r}_2^{2}}$. $Z_1$ and $Z_2$ are the effective charge of two nuclei of the molecule. Here, $\eta=0.5$ is the smoothing parameter which is used to avoid the Coulomb singularity. We take the center of mass as the coordinate origin. The relative location of the electron to the nuclear can be written as $\mathbf{r}_{1(2)}^2={\vert\mathbf{r}-\mathbf{R}_{1(2)}\vert}^2=[x\pm {R}_{1(2)}\cos\theta']^2+[y\pm {R}_{1(2)}\sin\theta']^2$, with $\mathbf{R}_{1(2)}$ is the vector from the coordinate origin to each nucleus, and ${R}_{1(2)}=Z_{2(1)}R/(Z_1+Z_2)$. $\theta'$ is the alignment angle (the angle between the molecular axis and the laser fundamental field), and we only consider the parallel alignment with $\theta'=0^{\circ}$. In numerical calculations, to explore the effects of the ionization potential $I_p$, we simulate two kinds of model polar molecules with different $I_p$ at different internuclear distances. To explore the effects of Coulomb potential, we also compare polar molecules to model symmetric molecules and atoms with similar $I_p$ to polar molecules. The specific parameters used in simulations are as follows. For the first kind of polar molecule, we choose the HeH$^{+}$ system. In this case, we fix $Z_1=1.6$ and $Z_2=0.8$. When the internuclear distance is $R=2$ a.u., the ionization potential of the ground state (the first excited state) is $I_{p0}=1.44$ a.u. ($I_{p1}=0.894$ a.u.). We mention that because of the importance of the excited state in strong-field ionization of polar molecules, besides of the ground state, we also show the calculated ionization potential of the excited state here. When $R=1.4$ a.u., the ionization potentials satisfy $I_{p0}=1.56$ a.u. and $I_{p1}=0.88$ a.u.. When $R=1.8$ a.u., the ionization potentials agree with $I_{p0}=1.48$ a.u. and $I_{p1}=0.89$ a.u.. For the second kind of polar molecule, we fix $Z_1=1.3$ and $Z_2=0.65$. When $R=2$ a.u., the ionization potentials satisfy $I_{p0}=1.11$ a.u. and $I_{p1}=0.645$ a.u.. When $R=1.4$ a.u., the ionization potentials are $I_{p0}=1.19$ a.u. and $I_{p1}=0.63$ a.u.. When $R=1.8$ a.u., the ionization potentials agree with $I_{p0}=1.14$ a.u. and $I_{p1}=0.64$ a.u.. For the first kind of symmetric molecule, we fix $Z_1=Z_2=1.24$ and $R=2$ a.u., and the ground-state ionization potential calculated satisfies $I_{p0}=1.44$ a.u.. For the second one, we fix $Z_1=Z_2=1.0$ and $R=2$ a.u., and the ionization potential calculated agrees with $I_{p0}=1.11$ a.u.. For atoms, we use the Coulomb potential having the form of $V(\mathbf{r})=-Z/\sqrt{\eta+\mathbf{r}^2}$, in which $Z$ is the whole effective charge and $\mathbf{r}^2=x^2+y^2$. For the first kind of atom with $Z=2.07$, the ground-state ionization potential calculated is $I_{p0}=1.44$ a.u.. For the second kind of atom with $Z=1.7$, the ionization potential calculated satisfies $I_{p0}=1.11$ a.u..

The term $\mathbf{E}(t)$ in Eq. (1) denotes the electric field of the laser. In elliptically-polarized cases, the electric field $\mathbf{E}(t)$ has the form of $\mathbf{E}(t)=f(t)[\vec{\mathbf{e}}_{x}E_{x}(t)+\vec{\mathbf{e}}_{y}E_{y}(t)]$, with $E_{x}(t)={E_0}\sin(\omega t)$ and $E_{y}(t)={E_1}\cos(\omega t)$, ${E_0}={E_L}/{\sqrt{1+\varepsilon^2}}$ and ${E_1}=\varepsilon{E_L}/{\sqrt{1+\varepsilon^2}}$. Here, $E_L$ is the  maximal laser amplitude corresponding to the peak intensity $I$, $\omega$ is the laser frequency, and $f(t)$ is the envelope function. The term $\varepsilon=0.87$ is the ellipticity, implying that the fundamental field $E_{x}(t)$ dominates in ionization. The term $\vec{\mathbf{e}}_{x}$($\vec{\mathbf{e}}_{y}$) is the unit vector along the $x(y)$ axis. We assume that the molecular axes of both asymmetric and symmetric molecules are oriented parallel to $\vec{\mathbf{e}}_{x}$, and for asymmetric molecules, the heavy (light) nucleus is located on the right (left) side. With the above treatments, we are actually considering the situation of perfect alignment and perfect orientation. We use trapezoidally shaped laser pulses with a total duration of fifteen cycles, which are linearly turned on and off for three optical cycles, and then kept at a constant intensity for nine additional cycles. The TDSE of $i\dot{\Psi}(\textbf{r},t)=$H$(t)\Psi(\textbf{r},t)$ is solved numerically using the spectral method \cite{Feit} with a time step of $\triangle t=0.05$ a.u.. We have used a grid size of $L_x\times L_y=409.6\times 409.6$ a.u. with space steps of $\triangle x=\triangle y=0.4$ a.u.. The numerical convergence is checked by using a finer grid. In addition, for different laser parameters used in numerical calculations of this paper, the Keldysh parameters $\gamma=\omega\sqrt{2I_{p0}}/E_0$ \cite{Keldysh1965} are all smaller than 1, indicating that we mainly discuss the case of tunneling ionization.

In order to avoid the reflection of the electron wave packet from the boundary and obtain the momentum space wave function, the coordinate
space is split into the inner and the outer regions with
${\Psi}(\textbf{r},t)={\Psi}_{in}(\textbf{r},t)+{\Psi}_{out}(\textbf{r},t)$, by multiplication using a mask function
$F(\mathbf{r})$. The mask function has the form of
$F(\mathbf{r})=F(x,y)=\cos^{1/2}[\pi(r_b-r_f)/(L_r-2r_f)]$ for $r_b\geq r_f$ and $F(x,y)=1$  for $r_b< r_f$.
Here, $r_b=\sqrt{x^2+y^2/\varepsilon^2}$, $r_f=2.1x_q$ with $x_q=E_0/\omega^2$  and $L_r/2=r_f+50$ a.u. with $L_r\leq L_x$.
In the inner region, the wave function ${\Psi}_{in}(\textbf{r},t)$ is propagated
with the complete Hamiltonian $H(t)$. In the outer region, the time evolution of the wave function ${\Psi}_{out}(\textbf{r},t)$ is carried out
in momentum space with the Hamiltonian of the free electron in the laser field.
The mask function is applied at each time  interval  of 0.5 a.u. and the obtained new fractions of the outer wave function are added to the momentum-space wave function $\tilde{{\Psi}}_{out}(\textbf{r},t)$ from which we obtain the PMD.
Then we find the local maxima of the PMD and the offset angle $\theta$ is obtained with a Gaussian fit of the angle distribution of local maxima.

\subsection{Analytical methods}
To analytically study the ionization of polar molecules in strong elliptical laser fields, we generalize the so-called tunneling-response-classical-motion (TRCM) model \cite{chen2021} to polar molecules. The TRCM model is a semiclassical model. It is based on SFA \cite{Lewenstein1995} but considers the effect of the Coulomb potential on tunneling ionization \cite{MishaY,Goreslavski,yantm2010}. Different from previous treatments, this model approaches the Coulomb effect from the perspective of Coulomb-related symmetry, allowing for an analytical description of the Coulomb effect. It is first proposed for describing tunneling ionization of atoms \cite{che2023}. Then it is generalized to symmetric molecules with small \cite{gen2022} and large \cite{shen2024} internuclear distances. This model can quantitatively reproduce recent attoclock experimental curves for atoms and provide a consistent explanation for relevant experimental phenomena \cite{chen2021,cheaps}. In this paper, we further develop the TRCM to describe strong-field ionization of polar molecules by introducing the PD effect. In \cite{WS2020}, a PD-included Coulomb-modified SFA (MSFA) model for polar molecules has been developed. The PD-included MSFA has also been used to study the effect of PD on Coulomb induced ionization time lag for polar molecules \cite{CheNJP2023}. However, the MSFA considers the Coulomb effect through numerical solution of Newton equation including both electric force and Coulomb force, so the detailed mechanism of how Coulomb potential affects tunneling ionization is still unclear and an analytical expression for this lag cannot be obtained.

\subsubsection{SFA for polar molecules}
\textit{Electron trajectories}. According to the SFA, tunneling ionization of atoms and molecules in strong laser fields can be described by electron trajectory, characterized by the photoelectron momentum $\textbf{p}$ and the time $t_s$ at which the tunneling event occurs \cite{Lewenstein1995}. For the polar molecule with a permanent dipole  $\mathbf{D}$ \cite{KamtaGL2005,Bandrauk903} which is directing from the heavy nucleus to the light nucleus, the electron trajectory ($\textbf{p},t_s$) agrees with the following saddle-point equation \cite{Madsen2010} 
\begin{equation}
[\textbf{p}+\textbf{A}(t_s)]^2/2=-I_{p0}-\mathbf{E}(t_s)\cdot \mathbf{D}.
\end{equation}
Here, $I_{p0}$ is the ionization potential of the ground state $\vert0\rangle$ of the system, and $\textbf{A}(t)$ is the vector potential of the electric field $\textbf{E}(t)$. The solution $t_s=t_0+it_x$ of the above equation is complex. The real part $t_0$ can be understood as the tunneling-out time. In SFA without considering the Coulomb effect, the tunneling-out time is also the ionization time. The momentum-time pair ($\textbf{p},t_0$) is also called the electron trajectory.

\textit{Effects of PD}. The expression of the PD used here has the from of $\mathbf{D}=\mathbf{D}_0-\mathbf{D}_1$ with $\mathbf{D}_0=\left\langle 0|\hat{\mathbf{r}}|0\right\rangle$ and $\mathbf{D}_1=\left\langle 1|\hat{\mathbf{r}}|1\right\rangle$. Here, the term $\vert1\rangle$ denotes the first excited state of the system. The terms $\mathbf{D}_0$ and $\mathbf{D}_1$ denote the PD related to the ground state $\vert0\rangle$ and the first excited state $\vert1\rangle$.  The expression of $\mathbf{D}=\mathbf{D}_0-\mathbf{D}_1$ considers the strong coupling between $\vert0\rangle$ and $\vert1\rangle$ \cite{Shang2017,Bandrauk903}. When the values of $\mathbf{D}_0$ and $\mathbf{D}_1$ depend on the selection of the coordinate origin, the value of $\mathbf{D}$ is independent of the selection. This expression of $\mathbf{D}=\mathbf{D}_0-\mathbf{D}_1$ has been used to explain complex strong-field phenomena related to asymmetric ionization \cite{WS2020,Che2021pra} and odd-even HHG \cite{ChenPRA2024} from polar molecules. One can see from Eq. (2), the interaction of the laser field and the PD will cause a time-dependent dressed ground-state energy $[-I_{p0}-\mathbf{E}(t)\cdot \textbf{D}]$. This will further lead to asymmetric ionization of the system in the first and second half cycles of the laser. When the ground-state energy is dressed upward, the ionization will be easier to occur. The situation reverses when the energy is dressed downward.

Once the trajectory ($\textbf{p},t_0$) is obtained, the corresponding amplitude $c(\textbf{p},t_0)$ for the trajectory can be evaluated with $c(\textbf{p},t_0)\sim e^{b}$. Here, $b$ is the imaginary part of the quasiclassical action 
\begin{equation}
S(\textbf{p},t_s)=\int_{t_s}\{{[\textbf{p}+\textbf{A}(t'})]^2/2+I'_p\}dt',
\end{equation}
Here, the term $I'_p=I_{p0}+\mathbf{E}(t')\cdot \mathbf{D}$ corresponds to the laser-dressed ground-state ionization potential. In this paper, we only consider the cases of the parallel alignment, namely, the molecular axis is aligned along the the main axis of the polarization ellipse. We also assume perfect alignment and perfect orientation of the polar molecule. As in TDSE, when considering that the heavy (light) nucleus of the polar molecule is located on the right (left) of the $x$ axis, we have $\mathbf{D}=D_x\vec{\mathbf{e}}_{x}$ with $D_x<0$. Then at the peak time $t_0$ of the laser field in the first (second) half of the laser cycle, we have $\mathbf{E}(t_0)=E_0\vec{\mathbf{e}}_{x}$ ($\mathbf{E}(t_0)=-E_0\vec{\mathbf{e}}_{x}$) and $I'_p=I_{p0}+E_0\cdot D_x$ ($I'_p=I_{p0}-E_0\cdot D_x$). The expressions show that for the present coordinate system, in the first (second) half laser cycle, the dressed ground-state ionization potential $I'_p$ is smaller (larger) than $I_{p0}$ and accordingly the ionization is easier (more difficult) to occur. The values of $\textit{D}=D_x$ for polar molecules with different internuclear distances $R$ used in our simulations are marked in Fig. 2 and Fig. 4 below. It should be mentioned that when the value of $D$ in Eqs. (2) and (3) equals zero, we return to the general SFA and Eqs. (2) and (3) can be used to evaluate electron trajectories for atoms and symmetric molecules. In addition, in \cite{Shang2017}, it has been shown that for longer laser wavelengths, the laser-dressed ground-state ionization channel dominates in ionization. However, for shorter laser wavelengths, the dressed-up ground state can strongly couple with the dressed-down first excited state, so the excited-state ionization channel may play an important role in ionization. In the paper, we mainly focus on the long-wavelength case.

\textit{Mapping between time and momentum}. According to electron trajectories in SFA, the mapping relation between the ionization time and the drift momentum can be written as 
\begin{equation}
\mathbf{p}=\textbf{v}(t_{0})-\textbf{A}(t_{0}).
\end{equation}
The term $\textbf{v}(t_{0})=\mathbf{p}+\textbf{A}(t_{0})$ denotes the exit velocity of the photoelectron at the tunnel exit $\mathbf{r}_0\equiv\mathbf{r}(t_0)=Re(\int^{t_0}_{t_s}[\mathbf{p}+\mathbf{A}(t')]dt')$ \cite{yantm2010}.

\subsubsection{TRCM for polar molecules}
\textit{Coulomb-included mapping between time and momentum}. Next, based on the frame of the TRCM model for atoms and symmetric molecules, we include the Coulomb effect into the SFA for polar molecules. The TRCM assumes \cite{chen2021,che2023} that when the Coulomb potential is considered, the tunneling electron with the drift momentum $\textbf{p}$ is still located at a quasi-bound state at the tunnel exit $\mathbf{r}(t_0)$, which approximately satisfies the virial theorem. This basic assumption related to the virial theorem in TRCM has been validated in recent theory studies \cite{chen2025,Wu2025}. Specifically, the average potential energy of the quasi-bound state is $\langle V(\mathbf{r})\rangle\approx V(\textbf{r}(t_0))$ and the average kinetic energy is $\langle\textbf{v}^2/2\rangle=n_f\langle \textbf{v}_x^2/2\rangle\approx-V(\textbf{r}(t_0))/2$. Here, the term $n_f=2,3$ is the dimension of the system studied. This quasi-bound state can be further approximately treated as a quasi-particle with the velocity $\textbf{v}_{i}$. This velocity  has the amplitude $v_i=|\textbf{v}_{i}|\approx\sqrt{|V({\textbf{r}}(t_0))|/n_f}$ and the direction of the velocity $\textbf{v}_{i}$ is opposite to the direction of the position vector $\textbf{r}(t_0)$ and points to the nucleus. That is $\textbf{v}_{i}=-v_i {\textbf{r}}(t_0)/|{\textbf{r}}(t_0)|$. This velocity reflects the basic symmetry requirement of the Coulomb potential on the electron wave packet at the tunnel exit and prevents the tunneling electron from escaping at the tunneling-out time $t_0$. Therefore, a small period of time $\tau$ is needed for the tunneling electron to obtain an impulse from the laser field to overcome this Coulomb-induced velocity $\textbf{v}_{i}$, namely $\int_{t_0}^{t_0+\tau} \textbf{E}(t) dt=\textbf{v}_i$. When the time $\tau$ is small enough, we have $|\textbf{E}(t_0)|\tau\approx|\textbf{v}_{i}|$. Then at the time $t_i=t_0+\tau$, the electron is free, and thereafter, the Coulomb potential can be neglected. With the discussions, the mapping relation between the ionization time $t_i$ and the Coulomb-included drift momentum $\mathbf{p}'$ in TRCM can be written as \cite{chen2021,che2023}
\begin{equation}
\mathbf{p}'=\textbf{v}(t_{0})+\textbf{v}_{i}-\textbf{A}(t_{0})\approx\textbf{v}(t_{0})-\textbf{A}(t_{i}).
\end{equation}
The above expression shows that the time $\tau=t_i-t_0$ manifests itself as the Coulomb-induced ionization time lag relative to the Coulomb-free ionization time $t_0$ in strong-field tunneling ionization. From a semiclassical perspective, the lag $\tau$ can be understood as the response time of the electron within an atom or a molecule to light in laser-induced photoelectric effects. From a quantum perspective, the time $\tau$ can also be understood as the response time of the electron wave function to a strong-field ionization event. We will discuss the specific expression of $\tau$ for different gas targets of atoms and molecules later.

\textit{Mapping between time and offset angle}. In attoclock related to ionization of atoms and molecules in strong near-circular laser fields, the offset angle in PMD is generally used as the characteristic quantity to extract electron ultrafast dynamics information. The offset angle $\theta$ is related to the most probable route (MPR), for which the tunnel event occurs at the peak time $t_0$ of the laser field with $|E_x(t_0)| = E_0$ and $v_x(t_0)=0$. According to Eq. (5), the angle $\theta$ of MPR satisfies the following relation 
\begin{equation}
\tan\theta=p'_x/p'_y= A_x(t_i)/[A_y(t_i)-v_y(t_0)].
\end{equation}
The above expression shows the inherent relation between the offset angle which can be observed in experiments and the ionization time which cannot be measured directly. The above relation can be further simplified when  the Keldysh parameter $\gamma=\omega\sqrt{2I_{p0}}/E_0$  \cite{Keldysh1965} is smaller than the unity, namely $\gamma\ll1$. In this case, the velocity $v_y(t_0)$ can be neglected, and we have $\tan\theta\approx A_x(t_i)/A_y(t_i)$. The expression has been called the adiabatic version of Eq. (6). Considering $t_i=t_0+\tau$ and $\omega t_0 = \pi /2$ or $3 \pi /2$ for the MPR, this expression can be further approximated as $\tan\theta\approx \tan(\omega \tau)/\varepsilon\approx \theta\approx \omega\tau/\varepsilon$, which shows a clear one-to-one correspondence between the PMD offset angle $\theta$ and the response time $\tau$.

Equation (6) also indicates that if the response time $\tau$ is obtained through analytical methods, the offset angle $\theta$ can be derived from this formula. Comparing the derived angle with the angle measured in the experiment can verify the applicability of Eq. (6). Next, we discuss the analytical expression of the lag $\tau$ for the MPR.

\textit{Response time for atoms}. According to the discussions in Eq. (5), the response time $\tau$ for MPR can be evaluated with $E_0 \tau\approx |\textbf{v}_{i}|$. Considering that $v_i\approx\sqrt{|V({\textbf{r}}(t_0))|/n_f}$, we have 
\begin{equation}
\tau\approx\sqrt{|V({\textbf{r}}(t_0))|/n_f}/E_0.
\end{equation}
In single-active electron approximation, the potential  $V(\textbf{r})$ for a hydrogen-like atom at the exit position $r_0= |\textbf{r}(t_0)|=|Re(\int^{t_0}_{t_s}[\mathbf{p}+\mathbf{A}(t')]dt')|$ can be considered to has the form of $V(\textbf{r}(t_0))\equiv V(r_0)=-Z_a/r_{0}$. Here, $Z_a$ is the whole effective charge of the atom. The exit position $r_0$ is evaluated with the saddle points given by Eq. (2) at $D=0$. For comparisons with TDSE, $Z_a$ can be chosen as that used in simulations.

\textit{Response time for symmetric molecules}. In comparison with the atom, the symmetric molecule has the internuclear distance $R$. For the molecule with a relatively small $R$, theory studies have shown that the bound wave packet composed of coherent superposition of higher energy bound eigenstates of $H_0$ still approximately satisfies the virial theorem \cite{chen2025}. Therefore, at the tunnel exit $r_0$, the tunneling electron still can be considered to be located at a quasi-bound state that agrees with the virial theorem. This implies that the relation $E_0 \tau\approx|\textbf{v}_{i}|\approx\sqrt{|V({\textbf{r}}(t_0))|/n_f}$ for MPR still holds in the molecular case \cite{gen2022}. The main differences between the atom and the molecule are that they have different forms of Coulomb potential. For a diatomic molecule in a strong laser field, due to the two-center characteristic of the Coulomb potential,  the ground-state energy of the molecule will be dressed upward by the laser field and the tunnel exit of the molecule is closer to the nucleus than the atom with a similar ionization potential. This effect can be considered in TRCM by using a corrected tunnel exit with $r'_0=r_0-R$ \cite{gen2022}. With this correction, the response time for symmetric molecules can be evaluated through 
\begin{equation}
\tau\approx\sqrt{|V({\textbf{r}'}(t_0))|/n_f}/E_0.
\end{equation}
Here, $V(\textbf{r}'(t_0))\equiv V(r'_0)=-Z_m/r'_0$ with $r'_0=r_0-{R}$. The term $Z_m$ is the whole effective charge of the symmetric molecule and $r_0$ is the exit position as evaluated in Eq. (7). Once the value of response time $\tau$ is obtained, by substituting $\tau$ into Eq. (6), one can obtain the TRCM prediction of the offset angle $\theta$ for symmetric molecules. This corrected tunnel exit $r'_0$ considers the characteristics of the laser-dressed molecular Coulomb potential near the atomic nuclei and plays an important role in the calculation of the response time for symmetric molecules. We will have more discussions on this point later. 
\begin{figure}[t]
\begin{center}
\rotatebox{0}{\resizebox *{9.0cm}{4.4cm} {\includegraphics {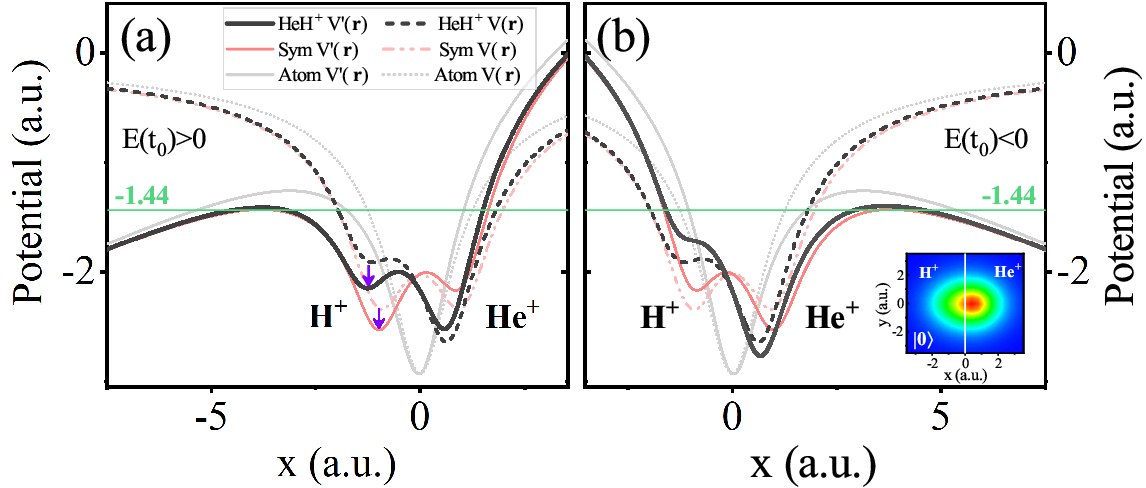}}}
\end{center}
\caption{(a): sketch of the laser-dressed Coulomb potential $V'(\textbf{r})=V(\textbf{r}) + E_{0} x$ (solid curves) and laser-free potential $V(\textbf{r})$ (dashed curves) for HeH$^{+}$ (black curves), symmetric molecule (red curves) and atom (gray curves) with similar ionization potentials for the first half cycle of laser fields at the peak time $t_0$ ($E(t_0)>0$). The horizontal green line indicates the ground state energy of $-I_{p0}=-1.44$ a.u.. (b): sketch for the second half cycle ($E(t_0)<0$) with $V'(\textbf{r})=V(\textbf{r}) - E_{0} x$. The laser amplitude $E_0$ used is $E_0=0.19$ a.u., and the internuclear distances of model molecules are all fixed at $R=2$ a.u.. The vertical purple arrows are used to guide the eyes. Embedded graph is the ground state wave function of HeH$^{+}$.
}
\label{fig:g1}
\end{figure}

\textit{Response time for polar molecules}. Next, we discuss the evaluation of the response time for polar molecules. Besides the PD induced energy shift as discussed in Eqs. (2) and (3), the structure of the laser-dressed asymmetric Coulomb potential near the atomic nuclei can also play a nontrivial role in strong-field ionization of the asymmetric molecule. In Fig. 1, we compare the laser-dressed potentials between atoms, symmetric and asymmetric molecules at the peak time $t_0$ of the laser field. One can observe from Fig. 1 (a) of $E(t_0)>0$, 1) the laser-Coulomb-formed barrier in the dressed potential for the atom is wider and higher than the barriers of symmetric and asymmetric molecules, which are similar to each other. 2) Around the atomic nuclei, the dressed potentials of atoms and molecules differ remarkably from each other. Due to the existence of the internuclear distance, the field-free molecular Coulomb potential show two potential wells near these two atomic nuclei, as shown by the black and red dashed lines. In strong laser fields, one potential well is dressed up and another is dressed down, as shown by the black and red solid curves. For the peak time of the laser field related to the MPR, the energy difference between the potential well with higher energy and that with lower energy is about $\Delta E=E_0R$. At intermediate internuclear distances, the electron is easier to escape from the upper potential well, inducing remarkable enhance of ionization yields which has been termed as charge-resonance enhanced ionization \cite{Bandrauk1995}. In \cite{gen2022}, it has been shown that by using the energy difference as a correction in the evaluation of the tunneling position $r_0'$ (namely $r'_0=r_0-\Delta E/E_0$ with $\Delta E=E_0R$ and $r_0$ being the tunnel exit evaluated with Eq. (2) at $D=0$), the developed TRCM is able to quantitatively reproduce the attoclock offset-angle results for aligned H$_2^+$ obtained in numerical experiments. For the present asymmetric case, besides the energy difference $E_0R$, the upper potential well is nearer to the laser-Coulomb-formed barrier through which the bound electron escapes, in comparison with the symmetric case, as indicated by the purple vertical arrows. This implies that when the direction of the laser polarization directs from the lighter nucleus to the heavier nucleus, the asymmetric molecule is easier to ionize than the symmetric molecule. Considering this characteristic, we further correct the exit position $r''_0$ by using the expression of $r''_0=r_0-\Delta E'/E_0$ with $\Delta E'=E_0R+E_0(R_2-R_1)$, $R_1=Z_2 R/(Z_1+Z_2)$ and $R_2=Z_1 R/(Z_1+Z_2)$. Here, $Z_1$ ($Z_2$) is the effective charge of the heavier (lighter) nucleus, and $r_0$ is the tunnel exit evaluated with Eq. (2) at $D\neq0$. This energy correction $E_0(R_2-R_1)$ in the above expression of $r''_0$ reflects the fact that due to the asymmetric Coulomb potential, excited states play a more important role in ionization of polar molecules than symmetric molecules, as discussed below Eq. (2). For $Z_1=Z_2$, we have $\Delta E'=\Delta E=E_0R$ and $D=0$. Then the expression of $r''_0$ returns to the symmetric case. We mention that this correction $E_0(R_2-R_1)$ is not applicable for cases of intermediate and large $R$. For these cases, due to the charge-resonance effect \cite{Bandrauk1995}, the ground state and the first excited state of the system will be strongly coupled together by the laser field and this strong coupling needs to be described fully quantum mechanics \cite{Chen2006,JPB45,shen2024}.

For the case of $E(t_0)<0$ in Fig. 1(b), one can observe that the structures of laser dressed Coulomb potentials for symmetric and asymmetric molecules around the atomic nucleus are comparable. In addition, the PD effect results in a larger energy difference between the ground state and excited states. That is, we can ignore the influence of excited states effect for $E(t_0)<0$. Then for the asymmetric molecule, the exit position $r''_0$ can be evaluated using $r''_0=r_0-\Delta E/E_0$ with $\Delta E=E_0R$ and $r_0$ being the tunneling position evaluated with Eq. (2) at $D\neq0$. We mention that for the symmetric molecule, the situations are the same for $E(t_0)>0$ and $E(t_0)<0$.

With the above discussions, the response time $\tau$ of MPR for polar molecules can be evaluated using 
\begin{equation}
\tau\approx\sqrt{|V({\textbf{r}''}(t_0))|/n_f}/E_0.
\end{equation}
Here, $V(\textbf{r}''(t_0))\equiv V(r''_0)=-Z_m/r''_0$ with $r''_0=r_0-{R}-(R_2-R_1)$ for $E(t_0)>0$ and $r''_0=r_0-{R}$ for $E(t_0)<0$. The term $Z_m$ is the whole effective charge of the asymmetric molecule. The above expressions of $r''_0$ take into account the characteristics of laser-dressed asymmetric Coulomb potential near the atomic nuclei and the influence of PD. Once the value of response time $\tau$ is obtained, by substituting $\tau$ into Eq. (6), one can obtain the TRCM prediction of the offset angle $\theta$ for polar molecules. In the paper, we will use Eq. (7) to calculate the response time $\tau$ for atoms, use Eq. (8) for symmetric molecules and Eq. (9) for polar molecules. Based on Eqs. (8) and (9), we can also analyze the difference of response time between symmetric and asymmetric molecules with similar ionization potentials. For the case of $E(t_0)>0$, due to the effect of PD, the laser dressed ionization potential of the asymmetric molecule is smaller than the symmetric one. In addition, due to the characteristics of the laser-dressed asymmetric Coulomb potential near the nuclei, the exit position for the asymmetric molecule is also nearer to the nuclei than the symmetric molecule. As a result, the response time $\tau$ is larger for the asymmetric molecule than the symmetric one. According to the relation $\theta\approx\omega\tau/ \varepsilon$, the offset angle related to MPR in the first half laser cycle for the asymmetric molecule is also larger than the symmetric one. For the case of $E(t_0)<0$, the laser-dressed potentials near the nuclei are comparable for symmetric and asymmetric molecules, but the PD effect will induce the laser-dressed ionization potential of the asymmetric molecule to be larger than the symmetric one. As a result, the response time and accordingly the offset angle for the asymmetric molecule is smaller than the symmetric one. In the following, through comparing the model predictions to TDSE, we will show that the TRCM model developed here for polar molecules can quantitatively describe the attoclock results of TDSE and the above analyses also hold in numerical experiments.

\textit{Coulomb-included PMDs}. Equations (7)-(9) are obtained for the specific electron trajectory of MPR related to the peak time of the laser field. They can also be extended to general SFA electron trajectories corresponding to arbitrary time $t_0$ with $\tau\approx\sqrt{|V({\textbf{r}}(t_0))|/n_f}/|\textbf{E}(t_0)|$ for atoms, $\tau\approx\sqrt{|V({\textbf{r}}'(t_0))|/n_f}/|\textbf{E}(t_0)|$ for symmetric molecules, and $\tau\approx\sqrt{|V({\textbf{r}}''(t_0))|/n_f}/|\textbf{E}(t_0)|$ for polar molecules. These terms $V({\textbf{r}}(t_0))$, $V({\textbf{r}}'(t_0))$ and $V({\textbf{r}}''(t_0))$ in the above expressions are as defined in Eqs. (7)-(9), but the exit position $r_0$ is evaluated at arbitrary time $t_0$. By further assuming that for an arbitrary SFA electron trajectory ($\textbf{p},t_0$), the Coulomb potential does not influence the corresponding complex amplitude  $c(\textbf{p},t_0)$, one can obtain the TRCM amplitude $c(\textbf{p}',t_i)$ for Coulomb-included electron trajectory ($\textbf{p}',t_i$) directly from the SFA one using the relation of $c(\textbf{p}',t_i)\equiv c(\textbf{p},t_0)\sim e^{b}$ and the mapping of $\mathbf{p}'=\textbf{v}(t_{0})-\textbf{A}(t_i=t_0+\tau)$. In the above expressions, $|\textbf{E}(t_0)|$ is the amplitude of the laser field. For the elliptical case, we have $|\textbf{E}(t_0)|=\sqrt{[E_0 \sin(\omega t_0)]^2+[E_1 \cos(\omega t_0)]^2}$. Then we can obtain the Coulomb-included PMD of $|c(\textbf{p}',t_i)|^2$.

\section{Results and discussion}
\begin{figure}[t]
\begin{center}
\rotatebox{0}{\resizebox *{7.5cm}{9.5cm} {\includegraphics {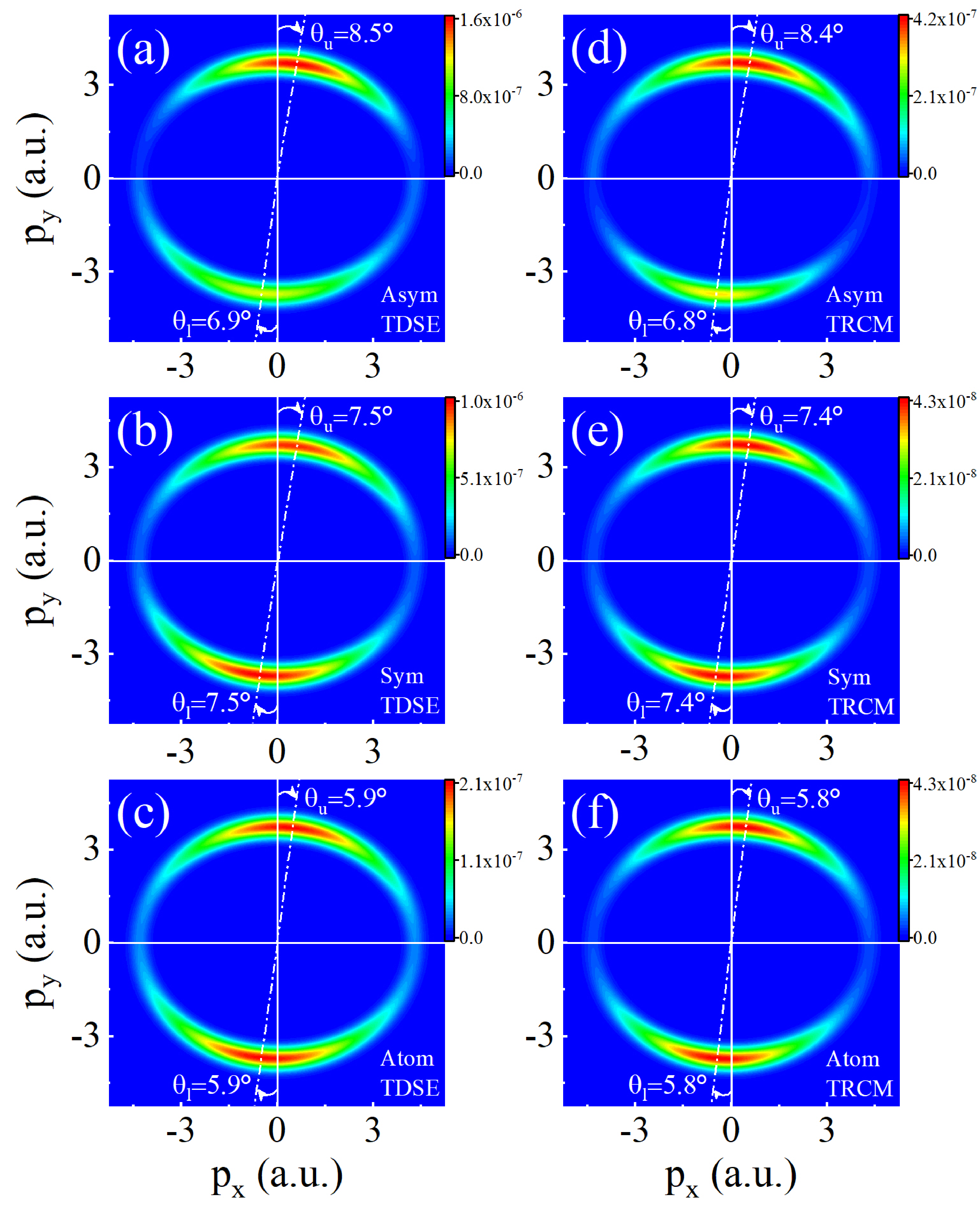}}}
\end{center}
\caption{PMDs of model molecules and atom obtained by TDSE (left column) and TRCM (right column). (a): PMD of HeH$^+$ obtained by TDSE (Asym TDSE) at $R=2$ a.u., with permanent dipole $D=-0.7$ a.u.. (b): PMD of symmetric molecule obtained by TDSE (Sym TDSE) at $R=2$ a.u.. (c): PMD of  atom obtained by TDSE (Atom TDSE). (d)-(f) are PMDs obtained by TRCM using the same laser and molecular parameters as TDSE. The offset angles $\theta$ are indicated by the white arrows. Model molecules and atom have the same ground state ionization potential $I_{p0}=1.44$ a.u.. Laser parameters used are $I=2.25\times10^{15}$ W/cm$^{2}$ (with $E_0=0.19$ a.u.), $\lambda=1000$ nm.}
\label{fig:g2}
\end{figure}

In Fig. 2, we show the PMDs of HeH$^+$, symmetric molecule and atom obtained by TDSE (left column) and TRCM (right column). The internuclear distances of molecules are fixed at $R=2$ a.u., and molecules and atom have the same ground state ionization potential $I_{p0}=1.44$ a.u.. Firstly, it can be seen from the PMD of HeH$^+$ in Fig. 2 (a) that the upper half plane is brighter than that of the lower half, indicating that ionization is stronger in the first half cycle. This phenomenon can be understood by the asymmetric ionization caused by the PD-induced Stark shift of the ground state along opposite directions during the first and second half cycles of the laser \cite{CheNJP2023}. In addition, the offset angle of the upper half plane ($\theta_{u}=8.5^{\circ}$) is larger than that of the lower half plane ($\theta_{l}=6.9^{\circ}$). The difference in offset angles can be understood by the difference in tunneling exit positions between adjacent laser sub-cycles. Specifically, the tunneling exit position of electrons is smaller (larger) in the first (second) half of the laser cycle, and the Coulomb force felt by electrons is stronger (weaker), corresponding to longer (shorter) response time and larger (smaller) offset angles. Figure 2(d) shows the PMD of HeH$^+$ calculated by TRCM. One can see that the result of TRCM not only reproduces the asymmetric ionization phenomenon in the upper and lower half planes of PMD, but also the offset angles are quantitatively consistent with TDSE ones. In calculation of Fig. 2 (d), we solve the saddle point Eq. (2), as well as the quasi-classical action Eq. (3) with $D=-0.7$ a.u.. Then, we obtain the response time $\tau\approx\sqrt{|V({\textbf{r}}''(t_0))|/n_f}/|\textbf{E}(t_0)|$, and draw PMD with the final momentum Eq. (5) and the amplitude $|c(\textbf{p}',t_i)|^2$. In the second and third rows of Fig. 2, we show the PMDs of symmetric molecule and atom. One can also see that TRCM results reproduce phenomena of TDSE, and the offset angles of TRCM are quantitatively consistent with TDSE ones. For symmetric molecule or atom, the Coulomb forces felt by electrons at tunnel exit positions in the first and the second half cycles are the same, so the offset angles of the upper ($\theta_{u}$) and lower ($\theta_{l}$) half planes of PMD are also the same. In calculations of Figs. 2(e) and 2(f), we solve Eqs. (2) and (3) with $D=0$, as well as the response time $\tau\approx\sqrt{|V({\textbf{r}}'(t_0))|/n_f}/|\textbf{E}(t_0)|$ for symmetric molecule and $\tau\approx\sqrt{|V({\textbf{r}}(t_0))|/n_f}/|\textbf{E}(t_0)|$ for atom, and draw PMD with the final momentum Eq. (5) and the amplitude $|c(\textbf{p}',t_i)|^2$. 

\begin{figure}[t]
\begin{center}
\rotatebox{0}{\resizebox *{9.0cm}{4.2cm} {\includegraphics {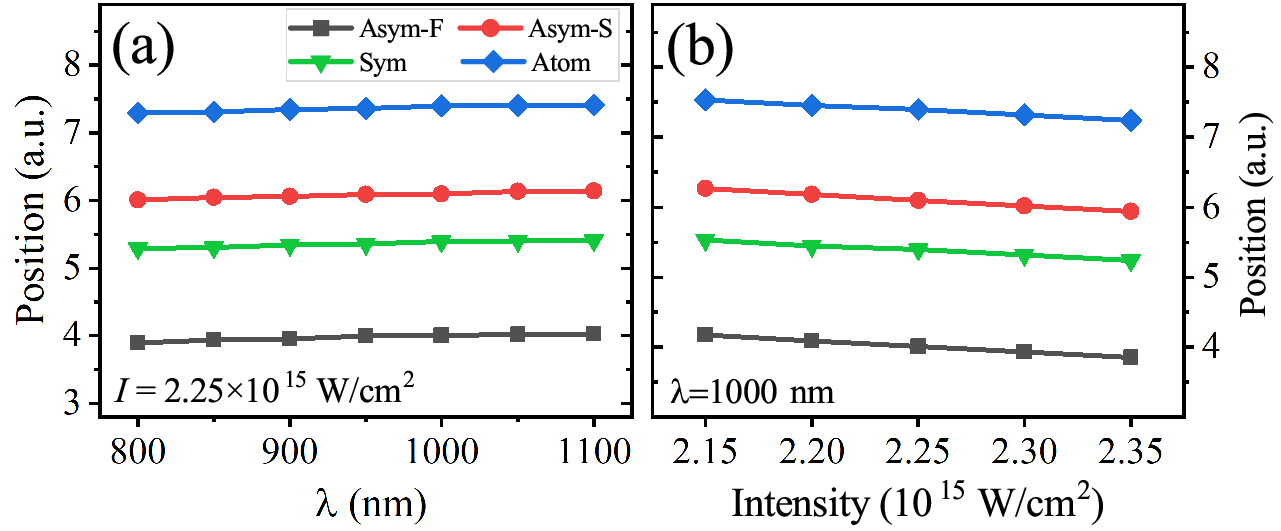}}}
\end{center}
\caption{Comparing the tunneling exit positions of asymmetric molecule, symmetric molecule, and atom at different laser parameters. The molecular and atomic parameters used are the same as in Fig. 2. For asymmetric molecule, the tunneling exit position $r''_0$ for the first (second) half of laser cycle is marked by ``Asym-F'' (``Asym-S''). The exit position for symmetric molecule $r'_0$ (atom $r_0$) is marked by ``Sym'' (``Atom''). The calculations for $r_0$, $r'_0$, and $r''_0$ can be found in discussions of Eqs. (7)-(9). In (a) [(b)], we change the laser wavelength (intensity), and fix the laser intensity (wavelength) at $I=2.25\times10^{15}$ W/cm$^{2}$ ($\lambda=1000$ nm).
}
\label{fig:g3}
\end{figure}

By carefully comparing PMD offset angles of HeH$^+$, symmetric molecule, and atom in Fig. 2, one can see that the larger angle of HeH$^+$ is greater than the angle of symmetric molecule, while the smaller angle is smaller than the angle of symmetric molecule but larger than the angle of atom. Here, to provide a more intuitive understanding for these phenomena, we compare the tunneling exit positions of asymmetric molecule, symmetric molecule, and atom with the same ground state ionization potential evaluated by TRCM in Fig. 3. The tunneling exit position for asymmetric molecule in the first (second) half cycle of the laser is calculated by $r''_0=r_0-{R}-(R_2-R_1)$ (by $r''_0=r_0-{R}$). The term $r_0$ is evaluated with Eq. (2) at $D=-0.7$ a.u.. For symmetric molecule (atom), the position is evaluated by $r'_0=r_0-{R}$ (by $r_0$), and $r_0$ being the tunnel exit evaluated with Eq. (2) at $D=0$. At the same laser parameters, the tunneling position of the asymmetric molecule in the first half cycle of the laser (black line) is the smallest, the position of the symmetric molecule (green line) is smaller than the second half cycle result of the asymmetric molecule (red line), and the position of the atom is the largest (blue line). According to Eqs. (7)-(9), one can obtain that the response time of MPR for atom satisfies $\tau\sim |V({\textbf{r}}(t_0))|\equiv Z_a/r_{0}$, for symmetric molecule $\tau\sim |V({\textbf{r}'}(t_0))|\equiv Z_m/r'_0$, and for asymmetric molecule $\tau\sim |V({\textbf{r}''}(t_0))|\equiv Z_m/r''_0$. When the ground state ionization potential is the same, the total effective charges $Z_m$ of symmetric and asymmetric molecules are comparable, and both are greater than that of atom $Z_a$. Therefore, at the same laser parameters, the response time $\tau$ for the asymmetric molecule in the first half cycle is the largest, the response time for the symmetric molecule is larger than the second half cycle result of the asymmetric molecule, and the response time of the atom is the smallest. According to the relation $\theta\approx\omega\tau/ \varepsilon$, one can obtain that the offset angle of the asymmetric molecule in the first half cycle is the largest, the angle of the symmetric molecule is larger than the second half cycle result of the asymmetric molecule, and this angle for the atom is the smallest.

\begin{figure}[t]
\begin{center}
\rotatebox{0}{\resizebox *{7.5cm}{9.5cm} {\includegraphics {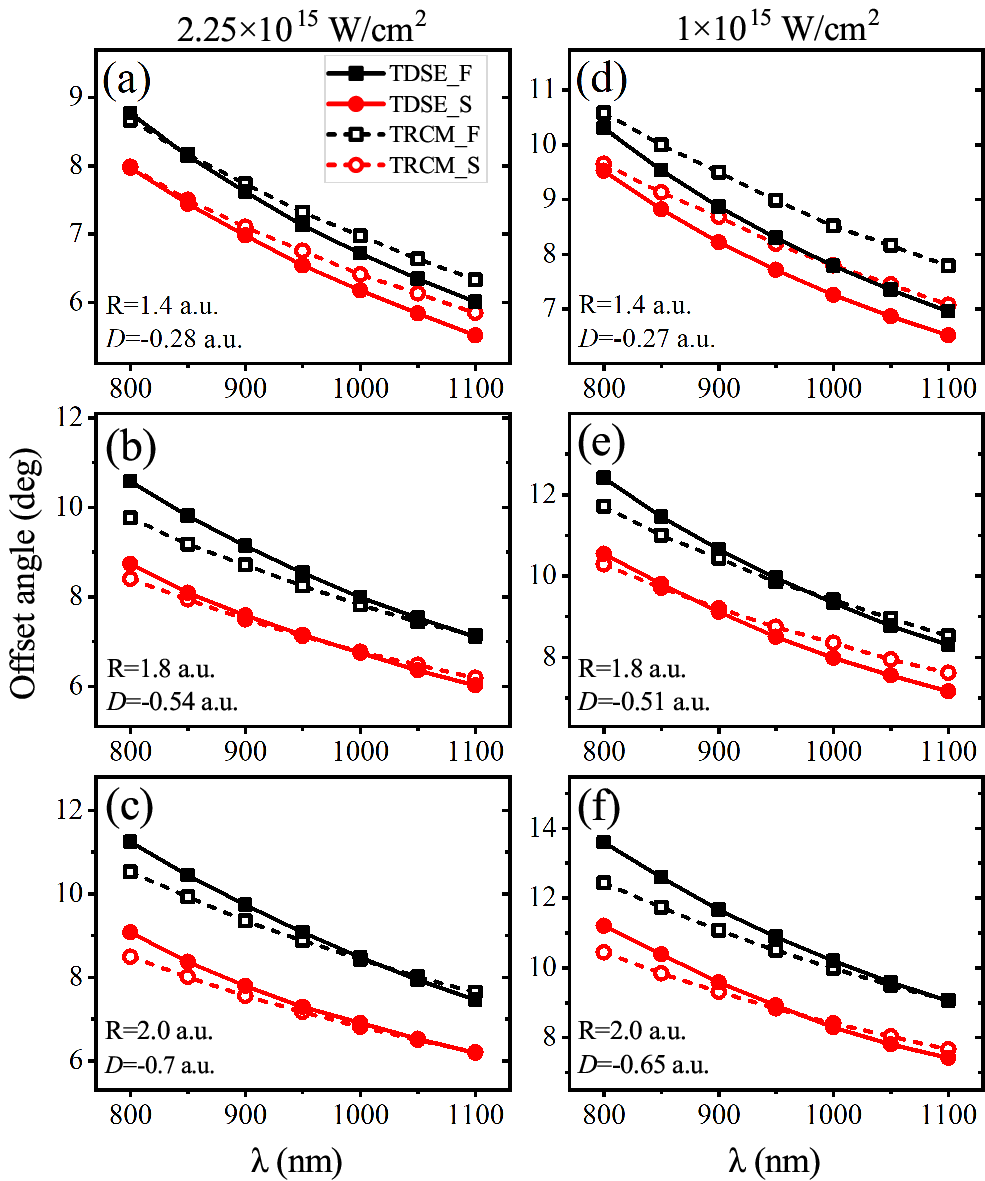}}}
\end{center}
\caption{Comparing the PMD offset angles of asymmetric molecules obtained by TDSE and TRCM. The black (red) solid line marked ``TDSE\_F(S)'' corresponds to the TDSE offset angles in the first (second) half cycle of the laser. The black (red) dashed line marked ``TRCM\_F(S)'' corresponds to the TRCM results in the first (second) half cycle. The asymmetric molecules studied in the left (right) column have a ground state ionization potential $I_{p0}=1.44$ a.u. ($I_{p0}=1.11$ a.u.) at $R=2$ a.u., and the laser intensity used is $I=2.25\times10^{15}$ W/cm$^{2}$ ($I=1\times10^{15}$ W/cm$^{2}$). The values of the internuclear distances $R$ and permanent dipole $D$ are marked in the left corner of each figure.}
\label{fig:g4}
\end{figure}

To validate  the above results, we further study attoclock of asymmetric molecules at different laser and molecular parameters. We also compare the results of asymmetric molecules to the corresponding symmetric molecules and atoms with the similar ionization potential $I_{p0}$. Firstly, in Fig. 4, we compare the PMD offset angles of asymmetric molecules at different internuclear distances calculated by TDSE and TRCM.

The asymmetric molecules studied in the left column of Fig. 4 have a ground state ionization potential $I_{p0}=1.44$ a.u. at $R=2$ a.u.. In TDSE simulations, we fix the total effective charge and smoothing parameter, and only change the the internuclear distance $R$ in molecular model potential $V(\mathbf{r})$ to study PMD offset angles of asymmetric molecules at different internuclear distances. The laser intensity used is $I=2.25\times10^{15}$ W/cm$^{2}$.

Figure 4(a) shows the result of $R=1.4$ a.u. with the permanent dipole $D=-0.28$ a.u.. One can see that the offset angles obtained by TDSE in the first half cycle (black solid line) are larger than those in the second half cycle (red solid line) at different wavelengths. These results are consistent with the phenomenon shown in Fig. 2(a). The black (red) dashed line marked as TRCM\_F(S) shows the offset angles obtained by TRCM for the first (second) half of laser cycle. One can see that the results of TRCM not only reproduce the changing trend of TDSE, but also the offset angles of TRCM in both half cycles are quantitatively consistent with TDSE ones (with a difference of less than 0.3 degrees). These results of TRCM first solve Eqs. (2) and (3), and calculate the response time $\tau$ with Eq. (9), and then substitutes $\tau$ into Eq. (6) to obtain the angles. Besides, one can also see that the offset angles of both half cycles decrease with increasing wavelengths. This phenomenon can also be explained by TRCM. According to Eq. (9), the response time of MPR for asymmetric molecule can be expressed as $\tau\approx\sqrt{Z_m/[n_f E^2_0 (r_0-R^{*})]}$. The term $R^{*}=R+(R_2-R_1)$ for $E(t_0)>0$, and $R^{*}=R$ for $E(t_0)<0$. For MPR, when the Keldysh parameter $\gamma\ll1$, the term $r_0=|\mathbf{r}(t_0)|\approx x(t_0)\approx (I'_p/E_0)[1-\gamma^2/4]\approx I'_p/E_0$ \cite{chen2021}. At the peak time $t_0$ of the laser field, in the first (second) half of the laser cycle, the term $I'_p=I_{p0}+E_0\cdot D$ ($I'_p=I_{p0}-E_0\cdot D$). That is, the response time $\tau$ is insensitive to changes in laser wavelength. According to the relation $\theta\approx\omega\tau/ \varepsilon$, as the wavelength increases, the laser frequency $\omega$ will decrease, and the PMD offset angle $\theta$ will also decrease. Figures 4(b) and 4(c) show the results of $R=1.8$ a.u. and $R=2$ a.u.. One can see that the changing trend of offset angles with laser wavelengths is consistent with Fig. 4(a), and as the internuclear distance increases, the absolute value of the permanent dipole $D$ increases, and the difference of angles between the first and second half cycles increases. In addition, the offset angles obtained by TRCM in both half cycles can still be quantitatively compared with the results of TDSE (with a difference of less than 0.8 degrees). The calculations of TRCM\_F(S) are similar to those introduced in Fig. 4(a).

\begin{figure}[t]
\begin{center}
\rotatebox{0}{\resizebox *{7cm}{11cm} {\includegraphics {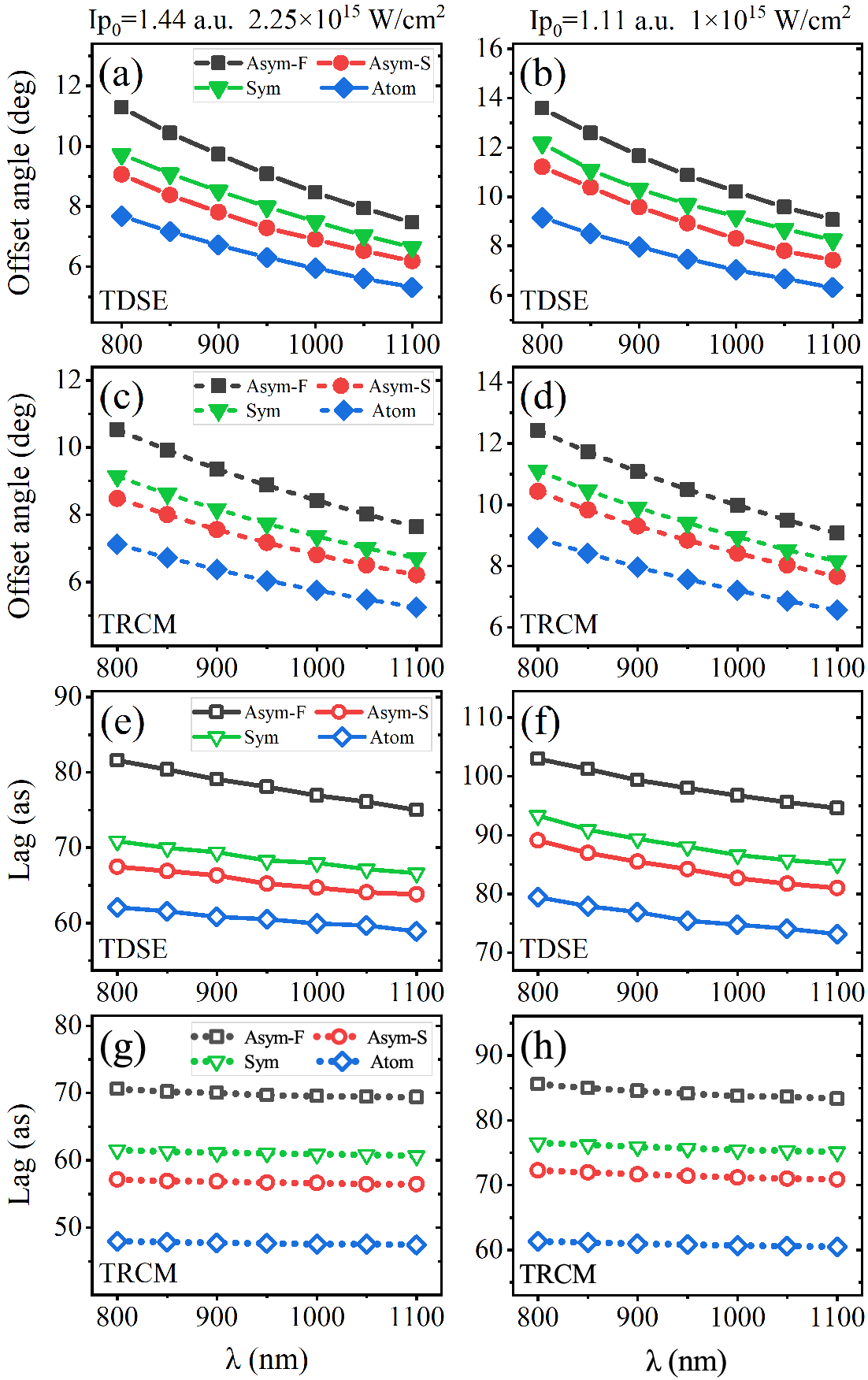}}}
\end{center}
\caption{Comparing the PMD offset angles and response time $\tau$ of asymmetric molecules, symmetric molecules, and atoms obtained by TDSE and TRCM at different laser wavelengths. The ground state ionization potentials of molecules and atoms in the left (right) column are fixed at $I_{p0}=1.44$ a.u. ($I_{p0}=1.11$ a.u.), and the laser intensity used is $I=2.25\times10^{15}$ W/cm$^{2}$ with $E_0=0.19$ a.u. ($I=1\times10^{15}$ W/cm$^{2}$ with $E_0=0.127$ a.u.). The internuclear distances of model molecules are all fixed at $R=2$ a.u.. The first (last) two rows show the comparisons of offset angles (response time $\tau$). The results of asymmetric molecules in the first (second) half of laser cycle are marked as ``Asym-F(S)''. The results of symmetric molecules (atoms) are marked as ``Sym'' (``Atom''). }
\label{fig:g5}
\end{figure} 

In the right column of Fig. 4, we show the results for another kind of asymmetric molecule which has a ground state ionization potential $I_{p0}=1.11$ a.u. at $R=2$ a.u.. The simulations of TDSE and calculations of TRCM\_F(S) in the right column are consistent with those introduced in the left column, and the laser intensity used is $I=1\times10^{15}$ W/cm$^{2}$. One can see that at the same $R$, the changing trend of offset angles in the right column of Fig. 4 obtained by TDSE and TRCM is consistent with the results in the left column. In Figs. 4(d)-4(f), the offset angles obtained by TRCM in both half cycles  can also be quantitatively compared with the results of TDSE (with a difference of less than 1.0 degrees). These results shown in Fig. 4 validate the applicability of the TRCM model for asymmetric molecules  introduced in Sec. II (B).

In Fig. 5, we compare the PMD offset angles and response times $\tau$ of asymmetric molecules, symmetric molecules, and atoms obtained by TDSE and TRCM at different laser wavelengths. To obtain the TRCM results of asymmetric molecules in the left (right) column, we first solve Eqs. (2) and (3) with $D=-0.7$ a.u. ($D=-0.65$ a.u.), and calculate the response time $\tau$ with Eq. (9). Then we substitute $\tau$ into Eq. (6) to obtain the offset angles. For symmetric molecules (atoms), we solve Eqs. (2) and (3) with $D=0$, and calculate the response time $\tau$ with Eq. (8) [Eq. (7)]. Then we substitute $\tau$ into Eq. (6) to obtain the angles.

The first row of Fig. 5 shows the offset angles of TDSE. One can see that as the laser wavelength increases, the offset angles of molecules and atoms decrease. Besides, at the same laser parameters, the angle of asymmetric molecule in the first half of laser cycle is the largest (black solid line), the angle of the symmetric molecule (green solid line) is greater than the second half cycle result of asymmetric molecule (red solid line), and the angle of the atom is the smallest (blue solid line). These phenomena are consistent with Fig. 2. The second row of Fig. 5 shows the angles calculated by TRCM. The results of TRCM not only reproduce the changing trend of TDSE, but also can quantitatively compare with the angles of TDSE (with a difference of less than 1.0 degrees).

The third row shows the response time $\tau$ obtained by TDSE. Unlike the offset angle, which is related to the PMD and can be directly measured in experiments, the response time is related to the instantaneous ionization property of the laser-driven system, which is not easy to probe in experiments. However, in TDSE simulations, this time can be evaluated by approximately calculating the instantaneous ionization rate. Specifically, we first find the time $t_i$ which corresponds to the maximal value of the instantaneous ionization rate $P(t)=dI(t)/dt$ \cite{xie2020}. Here, $I(t)=1-\sum_{m}|\langle m|\Psi(t)\rangle|^2$ is the instantaneous ionization yield, $|m\rangle$ is the bound eigenstate of the field-free Hamiltonian $H_0$, and $|\Psi(t)\rangle$ is the TDSE wave function. The upper limit $m_u$ of $m$ is determined with the eigenenergy $E_{m_u}$ of the $m_u$th eigenstate agreeing with the semiclassical analysis, that is, $|E_{m_u}|\approx |V(\textbf{r}(t_0))|$ \cite{che2023}. For example, for the model atom with $I_{p0}=1.44$ a.u., the term $|E_{m_u}|\approx Z_a/r_0 \approx Z_a E_0/I_{p0} =0.27$ a.u. and $m_u=8$. For the model atom with $I_{p0}=1.11$ a.u.,  $|E_{m_u}|\approx 0.19$ a.u. and $m_u=8$. Then the response time $\tau$ is obtained with $\tau=t_i-t_0$, where $t_0$ is the neighboring peak time of the laser field, agreeing with $|E(t_0)|=E_0$. One can see that as the wavelength increases, the TDSE response times show a similar changing trend to the offset angles in the first row. In addition, the response time of asymmetric molecule in the first half cycle is still the largest, the response time of symmetric molecule is greater than the second half cycle result of asymmetric molecule, and the result of atom is the smallest. These comparisons further demonstrate the well correlation between the PMD offset angle and response time $\tau$. The fourth row of Fig. 5 shows the response time obtained by TRCM. One can see that the results of TRCM are not sensitive to changes in laser wavelength and show a slight downward trend with increasing wavelength. These phenomena are consistent with our discussions in Fig. 4(a). The response times of TRCM can also be quantitatively compared with the results of TDSE (with a difference of less than 15 as).

\begin{figure}[t]
\begin{center}
\rotatebox{0}{\resizebox *{7cm}{11cm} {\includegraphics {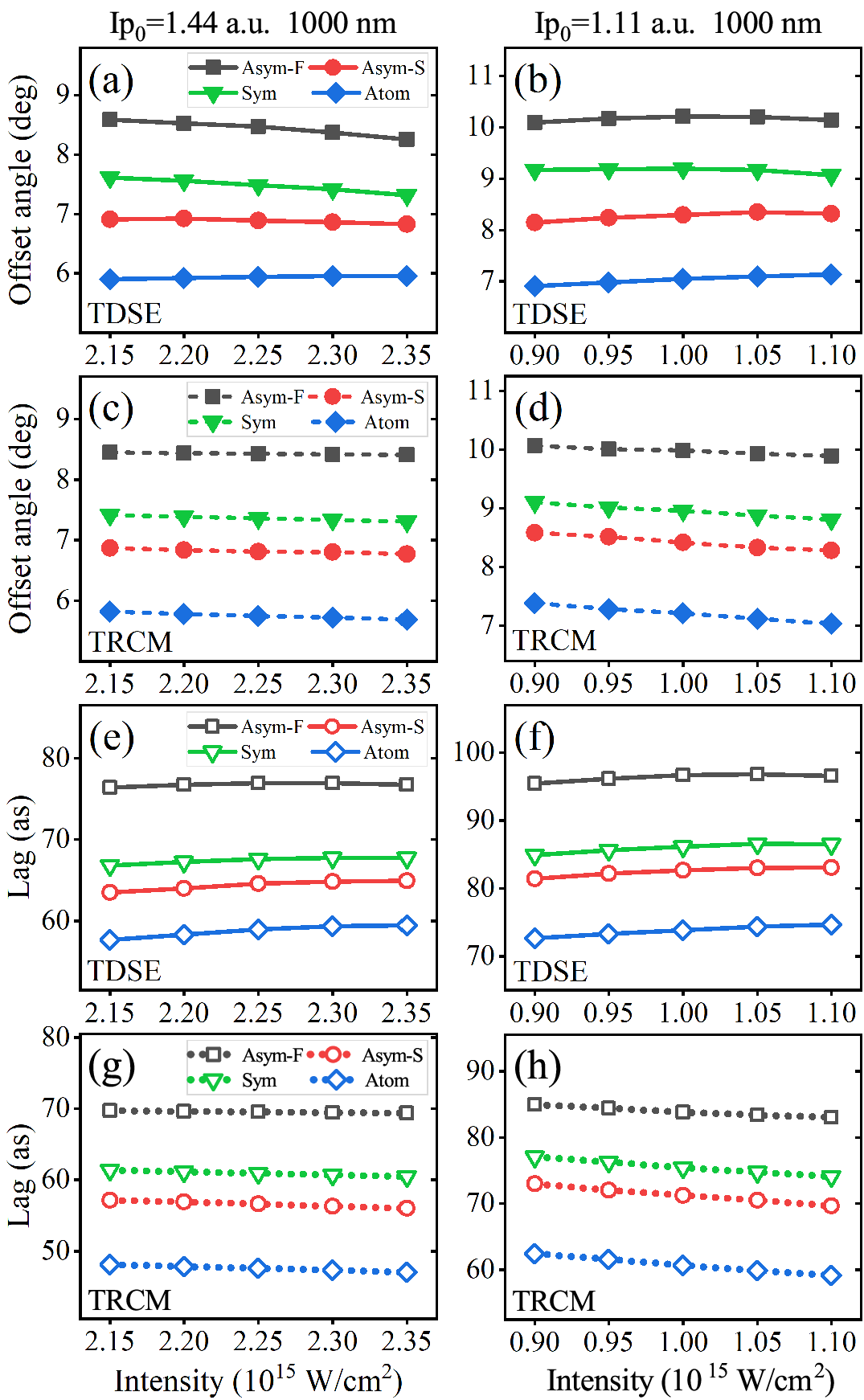}}}
\end{center}
\caption{Comparing the PMD offset angles and response time $\tau$ of asymmetric molecules, symmetric molecules, and atoms obtained by TDSE and TRCM at different laser intensities. The ground state ionization potentials of molecules and atoms in the left (right) column are fixed at $I_{p0}=1.44$ a.u. ($I_{p0}=1.11$ a.u.), the internuclear distances are all fixed at $R=2$ a.u., and the laser wavelength uses $1000$ nm. The meanings of different marked lines and the calculations for offset angle and $\tau$ of molecules and atoms are similar to those introduced in Fig. 5.
}
\label{fig:g6}
\end{figure} 
 
In Fig. 6, we further compare the PMD offset angles and response time $\tau$ of asymmetric molecules, symmetric molecules, and atoms obtained by TDSE and TRCM at different laser intensities. In the first row, one can see that the TDSE offset angles of model molecules and atoms are not sensitive to the change of the laser intensities for the relatively small parameter range explored here. This phenomenon can also be explained by TRCM. According to Eqs. (7)-(9), the response time $\tau$ of MPR for model molecules and atoms can be further simplified as $\tau\approx\sqrt{Z_s/[n_f E^2_0 (r_0-R_c)]}$. Here, the term $Z_s$ is the total effective charge of the system, and $R_c$ is a small correction parameter related to the internuclear distance $R$ and for atom $R_c=0$. For molecules with relatively small internuclear distance  $R$ and atoms, when the Keldysh parameter $\gamma\ll1$, the term $r_0\approx I_{p0}/E_0$. Then, the relation between response time $\tau$ and laser intensity can be simply approximated as $\tau\sim (E_0)^{-1/2}$. Therefore, the response time $\tau$ and offset angle $\theta\approx\omega\tau/ \varepsilon$ are not sensitive to the change of the laser intensities for the relatively small parameter range explored here. In addition, one can see that at the same laser parameters, the angle of asymmetric molecule in the first half of laser cycle is still the largest (black solid line), the angle of symmetric molecule (green solid line) is greater than the second half cycle result of asymmetric molecule (red solid line), and the angle of atom (blue solid line) is the smallest. In the second row, the offset angles of model molecules and atoms obtained by TRCM are also insensitive to the change of laser intensities, showing a slow decreasing trend with increasing intensity. Besides, the angles of TRCM can still be quantitatively compared with the results of TDSE (with a difference of less than 0.5 degrees). The calculations for offset angles and response time $\tau$ of TRCM in Fig. 6 are consistent with those introduced Fig. 5. In the third row, one can see that the TDSE response times of model molecules and atoms are also not sensitive to changes of laser intensities, and the relation between response times of molecules and atoms corresponds well to the offset angles in the first row. Here, the statistical method for TDSE response time is consistent with the method used in the third row of Fig. 5. The last row of Fig. 6 shows the comparisons of response time $\tau$ of model molecules and atoms obtained by TRCM. The TRCM results correspond well with the response times of TDSE and can quantitatively compare with TDSE ones (with a difference of less than 15 as).

\section{Conclusion} 
In summary, we have studied the tunneling ionization of oriented polar molecules in strong near-circular laser fields numerically and analytically. We focus on effects of the permanent dipole (PD) and the near-nucleus Coulomb potential of the polar molecule on the attoclock offset angle in photoelectron momentum distribution (PMD). We strive to provide an analytical expression for the response time of electrons in polar molecules to light in strong field tunneling ionization, through which the offset angle of TDSE can be analytically and quantitatively deduced. We have explored a wide range of laser and atomic parameters including different laser intensities, laser wavelengths and ionization potentials. We have also compared the results of polar molecules with those of symmetric molecules and atoms.

Our TDSE simulations show that for ionization events that occur in the first half of the laser cycle where the polarization direction of the laser field is antiparallel to the PD direction in our case, the PMD offset angle of polar molecules is greater than the offset angle associated with the ionization events that occur in the second half of the laser cycle where the polarization direction of the laser field is parallel to the PD direction. The angle difference between the first and second half cycles increases with the increase of internuclear distance. In comparison with model symmetric molecules and atoms with similar ionization potentials to polar molecules, the larger offset angle of polar molecules related to the first half laser cycle is larger than that of symmetric molecules and the smaller one related to the second half laser cycle is smaller than the angle of symmetric molecules but is larger than that of atoms. This phenomenon holds for different laser intensities and wavelengths, as well as polar molecules with different ionization potentials.

By using a developed strong-field model termed as TRCM which considers effects of both the PD and the characteristics of the asymmetric two-center Coulomb potential dressed by laser near the atomic nuclei, we are able to give an analytical expression for the response time and quantitatively reproduce the sub-cycle-related phenomena of the offset angle mentioned above. This model reveals that 1) the PD effect can increase (decrease) the response time of the tunneling electron inside polar molecules to light in photoemission and accordingly increase (decrease) the offset angle related to the first (second) half laser cycle. 2) For polar molecules, symmetric molecules and atoms with similar ionization potentials, the laser-dressed Coulomb potentials near the atomic nuclei differ remarkably from each other. Due to this difference, the tunnel exit of symmetric molecules is closer to the atomic nucleus than the tunnel exit of atoms. For ionization of polar molecules in the first half laser cycle, the excited state also plays a role, making the tunnel exit of polar molecules nearer to the nucleus than that of symmetric molecules. The situation reverses for ionization of polar molecules in the second half laser cycle. As a result, compared to the tunneling electrons of symmetric molecules, the tunneling electrons of polar molecules that escape in the first (second) half laser cycle experience stronger (weaker) Coulomb forces at the tunnel exit. Accordingly, the tunneling electrons of polar molecules require a longer (shorter) response time to overcome the Coulomb force that prevents them from escaping. The offset angle related to the response time is also larger (smaller) for polar molecules ionized in the first (second) half laser cycle than that for symmetric molecules. 

The model developed here may help to quantitatively study the attosecond-resolved ionization dynamics of polar molecules in strong laser fields. It is suitable for cases of longer laser wavelengths, in which the ground-state ionization channel dominates the ionization of polar molecules, and the role of excited states is relatively small. For cases of shorter laser wavelengths, the first excited state of polar molecules can play a more important role, and electrons can be directly ionized from the field-free excited state. In this case, a model that fully considers the contributions of the excited-state ionization channel to tunneling ionization needs to be developed.

\section{Acknowledgments}
This work was supported by the National Natural Science Foundation of China (Grant No. 12174239).


\begin{thebibliography}{2}
\bibitem{McPherson} A. McPherson, G. Gibson, H. Jara, U. Johann, T. S. Luk, I. A. McIntyre, K. Boyer, C. K. Rhodes, Studies of multiphoton production of vacuum-ultraviolet radiation in the rare gases, J. Opt. Soc. Am. B \textbf{4}, 595 (1987).

\bibitem{Schafer2} A. L'Huillier, K. J. Schafer, and K. C. Kulander, Theoretical aspects of intense field harmonic generation, J. Phys. B \textbf{24}, 3315 (1991). 


\bibitem{Tong1997} X. M. Tong and S. I. Chu, Theoretical study of multiple high-order harmonic generation by intense ultrashort pulsed laser fields: A new generalized pseudospectral time-dependent method, Chem. Phys. \textbf{217}, 119 (1997).


\bibitem{CorkumNP2007} P. B. Corkum and F. Krausz, Attosecond science, Nat. Phys. \textbf{3}, 381 (2007).


\bibitem{TaoZ2016} Z. Tao, C. Chen, T. Szilvasi, M. Keller, M. Mavrikakis, H. Kapteyn, and M. Murnane, Direct time-domain observation of attosecond final-state lifetimes in photoemission from solids, Science \textbf{353}, 62 (2016).


\bibitem{NeutzeR2000} R. Neutze, R. Wouts, D. van der Spoel, E. Weckert, and J. Hajdu, Potential for biomolecular imaging with femtosecond X-ray pulses, Nature (London) \textbf{406}, 752 (2000).



\bibitem{HentschelM2001} M. Hentschel, R. Kienberger, C. Spielmann, G. A. Reider, N. Milosevic, T. Brabec, P. Corkum, U. Heinzmann, M. Drescher, and F. Krausz, Attosecond metrology, Nature (London) \textbf{414}, 509 (2001).


\bibitem{UdemT2003} A. Baltuška, T. Udem, M. Uiberacker, M. Hentschel, E. Goulielmakis, C. Gohle, R. Holzwarth, V. S. Yakovlev, A. Scrinzi, T. W. Hänsch, and F. Krausz, Attosecond control of electronic processes by intense light fields, Nature (London) \textbf{421}, 611 (2003).



\bibitem{Schafer1} K. J. Schafer, B. Yang, L. I. DiMauro, and K. C. Kulander, Above threshold ionization beyond the high harmonic cutoff, Phys. Rev. Lett. \textbf{70}, 1599 (1993). 


\bibitem{Yang1993} B. Yang, K. J. Schafer, B. Walker, K. C. Kulander, P. Agostini, and L. F. DiMauro, Intensity-dependent scattering rings in high order above-threshold ionization, Phys. Rev. Lett. \textbf{71}, 3770 (1993).


\bibitem{Lewenstein1995} M. Lewenstein, K. C. Kulander, K. J. Schafer, and P. H. Bucksbaum, Rings in above-threshold ionization: A quasiclassical analysis, Phys. Rev. A \textbf{51}, 1495 (1995).


\bibitem{Paulus2002} W. Becker, F. Grasbon, R. Kopold, D. B. Milo\u{s}evi\'{c}, G. G. Paulus, and H. Walther, Above-threshold ionization: from classical features to quantum effects, Adv. At. Mol. Opt. Phys. \textbf{48}, 35 (2002). 


\bibitem{Corkum} P. B. Corkum, Plasma perspective on strong field multiphoton ionization, Phys. Rev. Lett. \textbf{71}, 1994 (1993).


\bibitem{Lewenstein1994} M. Lewenstein, Ph. Balcou, M. Yu. Ivanov, A. L'Huillier, and P. B. Corkum, Theory of high-harmonic generation by low-frequency laser fields, Phys. Rev. A \textbf{49}, 2117 (1994).




\bibitem{Ivanov1} Hiromichi Niikura, F. Legare, R. Hasbani, Misha Yu. Ivanov, D. M. Villeneuve, and P. B. Corkum, Probing molecular dynamics with attosecond resolution using correlated wave packet pairs, Nature \textbf{421}, 826 (2003). 


\bibitem{Corkum2} D. Zeidler, A. Staudte, A. B. Bardon, D. M. Villeneuve, R. D\"{o}rner, and P. B. Corkum, Controlling Attosecond Double Ionization Dynamics via Molecular Alignment, Phys. Rev. Lett. \textbf{95}, 203003 (2005). 

\bibitem{Becker1} W. Becker, X. Liu, P. J. Ho, and J. H. Eberly, Theories of photoelectron correlation in laser-driven multiple atomic ionization, Rev. Mod. Phys. \textbf{84}, 1011 (2012). 



\bibitem{Krausz2009} F. Krausz and M. Ivanov, Attosecond physics, Rev. Mod. Phys. \textbf{81}, 163 (2009).




\bibitem{Vrakking2014} F. L\'{e}pine, M. Y. Ivanov, and M. J. J. Vrakking, Attosecond molecular dynamics: fact or fiction? Nature Photon. \textbf{8}, 195 (2014).




\bibitem{Eckle2008} P. Eckle, A. N. Pfeiffer, C. Cirelli, A. Staudte, R. D\"{o}rner, H. G. Muller, M. Buttiker, and U. Keller, Attosecond Ionization and Tunneling Delay Time Measurements in Helium, Science \textbf{322}, 1525 (2008). 



\bibitem{Ivanov2012} D. Shafir, H. Soifer, B. D. Bruner, M. Dagan, Y. Mairesse, S. Patchkovskii, M. Yu. Ivanov, O. Smirnova, and N. Dudovich, Resolving the time when an electron exits a tunneling barrier, Nature (London) \textbf{485}, 343 (2012). 



 













\bibitem{xie2020} X. J. Xie, C. Chen, G. G. Xin, J. Liu, and Y. J. Chen, Coulomb-induced ionization time lag after electrons tunnel out of a barrier, Opt. Express \textbf{28}, 33228 (2020).



\bibitem{MishaY} T. Brabec, M. Yu. Ivanov, and P. B. Corkum, Coulomb focusing in intense field atomic processes, Phys. Rev. A \textbf{54}, R2551 (1996).


\bibitem{Goreslavski} S. P. Goreslavski, G. G. Paulus, S. V. Popruzhenko, and N. I. Shvetsov-Shilovski, Coulomb Asymmetry in Above-Threshold Ionization, Phys. Rev. Lett. \textbf{93}, 233002 (2004).


\bibitem{yantm2010} T. M. Yan, S. V. Popruzhenko, M. J. J. Vrakking, and D. Bauer, Low-energy structures in strong field ionization revealed by quantum orbits, Phys. Rev. Lett. \textbf{105}, 253002 (2010).



\bibitem{chen2021} J. Y. Che, C. Chen, W. Y. Li, S. Wang, X. J. Xie, J. Y. Huang, Y. G. Peng, G. G. Xin, and Y. J. Chen, Response time of photoemission at quantum-classic boundary, arXiv:2111.08491.
\bibitem{che2023} J. Y. Che, J. Y. Huang, F. B. Zhang, C. Chen, G. G. Xin, and Y. J. Chen, Roles of laser ellipticity in attoclocks, Phys. Rev. A \textbf{107}, 043109 (2023).
\bibitem{cheaps} J. Y. Che, C. Chen, W. Y. Li, W. Li, and Y. J. Chen,  Advances in response time of strong-field ionization of atoms, Acta Phys. Sin. \textbf{72}, 193301 (2023).

\bibitem{Sainadh2019} U. S. Sainadh, H. Xu, X. Wang, A. Atia-Tul-Noor, W. C. Wallace, N. Douguet, A. Bray, I. Ivanov, K. Bartschat, A. Kheifets, R. T. Sang, and I. V. Litvinyuk, Attosecond angular streaking and tunnelling time in atomic hydrogen, Nature \textbf{568}, 75 (2019).



\bibitem{Torlina2015} L. Torlina, F. Morales, J. Kaushal, I. Ivanov, A. Kheifets, A. Zielinski, A. Scrinzi, H. G. Muller, S. Sukiasyan, M. Ivanov, and O. Smirnova, Interpreting attoclock measurements of tunnelling times, Nat. Phys. \textbf{11}, 503-508 (2015).



\bibitem{Pfeiffer2012} A. N. Pfeiffer, C. Cirelli, M. Smolarski, D. Dimitrovski, M. Abu-samha, L. B. Madsen, and U. Keller, Attoclock reveals natural coordinates of the laser-induced tunnelling current flow in atoms, Nat. Phys. \textbf{8}, 76-80 (2012).



\bibitem{Boge2013} R. Boge, C. Cirelli, A. S. Landsman, S. Heuser, A. Ludwig, J. Maurer, M. Weger, L. Gallmann, and U. Keller, Probing Nonadiabatic Effects in Strong-Field Tunnel Ionization, Phys. Rev. Lett. \textbf{111}, 103003 (2013).


\bibitem{Quan2019} W. Quan, V. V. Serov, M. Z. Wei, M. Zhao, Y. Zhou, Y. L. Wang, X. Y. Lai, A. S. Kheifets, and X. J. Liu, Attosecond Molecular Angular Streaking with All-Ionic Fragments Detection, Phys. Rev. Lett. \textbf{123}, 223204 (2019).

\bibitem{Ramos2020} R. Ramos, D. Spierings, I. Racicot and A. M. Steinberg, Measurement of the time spent by a tunnelling atom within the barrier region. Nature \textbf{583}, 529 (2020).

\bibitem{Li2022} S. C. Li, A general scenario of tunneling time in different energy regimes, New J. Phys. \textbf{24}, 083033 (2022).


\bibitem{Wu2023} J. N. Wu, J. Y. Che, F. B. Zhang, C. Chen, W. Y. Li, G. G. Xin, and Y. J. Chen, Two-Color Attosecond Chronoscope, Opt. Express \textbf{31}, 21038 (2023).


\bibitem{gen2022} Y. G. Peng, J. Y. Che, F. B. Zhang, X. J. Xie, G. G. Xin, and Y. J. Chen, Response time of an electron inside a molecule to light in strong-field ionization, Opt. Express \textbf{32}, 12734 (2024).


\bibitem{shen2024} S. Q. Shen, Z. Y. Chen, S. Wang, J. Y. Che, and Y. J. Chen, Coulomb effects on strong-field ionization of stretched H2+, Phys. Rev. A \textbf{110}, 033106 (2024). 

\bibitem{Lein2002} M. Lein, N. Hay, R. Velotta, J. P. Marangos, and P. L. Knight. Interference effects in high-order harmonic generation with molecules, Phys. Rev. A \textbf{66}, 023805 (2002).
\bibitem{Itatani2004} J. Itatani, J. Levesque, D. Zeidler, Hiromichi Niikura, J. C. Kieffer, P. B. Corkum, and D. M. Villeneuve, Tomographic imaging of molecular orbitals, Nature\textbf{432}, 867 (2004).
\bibitem{chen2009} Y. J. Chen and Bambi Hu, Strong-field approximation for diatomic molecules: Comparison between the length gauge and the velocity gauge, Phys. Rev. A \textbf{80}, 033408 (2009).


\bibitem{peng2025} Y. G. Peng, J. Y. Che, R. H. Xu, S. Wang, X. J. Xie, and Y. J. Chen, A Coulomb-included model for high-order harmonic generation from atoms, unpublished.


\bibitem{KamtaGL2005} G. L. Kamta and A. D. Bandrauk, Phase dependence of enhanced ionization in asymmetric molecules, Phys. Rev. Lett. \textbf{94}, 203003 (2005).



\bibitem{WS2020} S. Wang, J. Y. Che, C. Chen, G. G. Xin, and Y. J. Chen, Tracing the origins of an asymmetric momentum distribution for polar molecules in strong linearly polarized laser fields, Phys. Rev. A \textbf{102}, 053103 (2020).



\bibitem{Che2021pra} J. Y. Che, C. Chen, S. Wang, G. G. Xin, and Y. J. Chen, Characterizing subcycle electron dynamics of polar molecules by asymmetry in photoelectron momentum distributions, Phys. Rev. A \textbf{104}, 063104 (2021).



\bibitem{Feit} M. D. Feit, J. A. Fleck, Jr., and A. Steiger, Solution of the Schr$\mathrm{\ddot{o}}$dinger Equation by a Spectral Method, J. Comput. Phys. \textbf{47}, 412 (1982).



\bibitem{Keldysh1965} L. V. Keldysh, Ionization in the field of a strong electromagnetic wave, Sov. Phys. JETP \textbf{20}, 1307 (1965).






\bibitem{CheNJP2023} J. Y. Che, C. Chen, S. Wang, G. G. Xin, and Y. J. Chen, Measuring ionization time lag of polar molecules with a calibrated attoclock, New J. Phys. \textbf{25}, 013016 (2023).



\bibitem{Bandrauk903} X. B. Bian and A. D. Bandrauk, Multichannel Molecular High-Order Harmonic Generation from Asymmetric Diatomic Molecules, Phys. Rev. Lett. \textbf{105}, 093903 (2010).


\bibitem{Madsen2010} D. Dimitrovski, C. P. J. Martiny, and L. B. Madsen, Strong-field ionization of polar molecules: Stark-shift-corrected strong-field approximation, Phys. Rev. A \textbf{82}, 053404 (2010).



\bibitem{Shang2017} S. Wang, J. Cai, and Y. J. Chen, Ionization dynamics of polar molecules in strong elliptical laser fields, Phys. Rev. A \textbf{96}, 043413 (2017).


\bibitem{ChenPRA2024} J. Y. Che, C. Y. Zhang, J. N. Wu, W. Y. Li, and Y. J. Chen, Time-resolved studies on odd-even harmonic emission from oriented HeH$^{+}$ in strong orthogonal two-color laser fields, Phys. Rev. A \textbf{110}, 063509 (2024).  


\bibitem{chen2025} Z. Y. Chen, S. Q. Shen, Y. P. Li, Z. Q. Yang, J. Y. Che, and Y. J. Chen, Coulomb-related symmetry in laser-induced tunneling ionization of atoms and Molecules, Phys. Rev. A \textbf{111}, 053118 (2025).
\bibitem{Wu2025} J. N. Wu, Z. Y. Chen, S. Q. Shen, S. Wang, J. Y. Che, and Y. J. Chen, Aspects of harmonic emission in the attoclock, Opt. Express \textbf{33}, 35679 (2025). 

\bibitem{Bandrauk1995} T. Zuo, and A. D. Bandrauk, Charge-resonance-enhanced ionization of diatomic molecular ions by intense lasers,
Phys. Rev. A \textbf{52}, R2511 (1995).



\bibitem{Chen2006} Y. J. Chen, J. Chen, and J. Liu, Charge-resonance effect on harmonic generation by symmetric diatomic molecular ions in intense laser fields, Phys. Rev. A \textbf{74}, 063405 (2006).


\bibitem{JPB45} Y. J. Chen and B. Zhang, Strong-field approximations for the orientation dependence of the total ionization of homonuclear diatomic molecules with different internuclear distances, J. Phys. B \textbf{45}, 215601 (2012).



\bibitem [*] {2} liwy157@126.com


\end{thebibliography}
\end{document}